\documentclass[prl,twocolumn,aps,showpacs,superscriptaddress]{revtex4-1}

\usepackage{amsmath}
\usepackage{amssymb}
\usepackage{hyperref}
\usepackage{graphicx}
\usepackage{subfigure}
\usepackage[usenames,dvipsnames]{color}

\begin{document}


\title{Coreless vortex dipoles and bubbles in phase-separated 
       binary condensates}

\author{S. Gautam} 
\affiliation{Physical Research Laboratory,
             Navarangpura, Ahmedabad - 380 009, India}
\author{P. Muruganandam} 
\affiliation{ School of Physics, 
              Bharathidasan University, Tiruchirapalli 620 024, 
              Tamil Nadu, India}
\author{D. Angom}
\affiliation{Physical Research Laboratory,
         Navarangpura, Ahmedabad - 380 009, India}


\date{\today}
\begin{abstract}
   Vortex dipoles are generated when an obstacle moves through a superfluid.
   In case of phase-separated binary condensates, with appropriate 
   interaction parameters in pan-cake shaped traps, we show that coreless 
   vortex dipoles are created when a Gaussian obstacle beam traverses across 
   them above a critical speed. As the obstacle passes through the inner 
   component, it carries along a bubble of the outer component. Using 
   Thomas-Fermi approximation, we show that phase-separated binary condensates
   can either support vortices with empty or filled cores. 
   For time dependent obstacle potentials, ramped down in the present case, 
   relative energy scales of the system influence the dynamical evolution of 
   the binary condensate.
\end{abstract}



\maketitle


  Vortices are among the most fundamental topological structures in fluid 
flows. These occur in a variety of fluids at different length scales and in 
different settings. In Bose-Einstein condensates (BECs), vortices carry 
integral angular momenta and serve as evidence of superfluidity. Formation and 
dynamics of solitary and multiple vortices in single species condensates is 
well studied. These vortices with empty cores and a phase singularity 
at the center of the cores are {\em normal vortices}, whose analogues are found
in variety of other fluids. An important development in vortex dynamics is the 
realization
of vortex dipoles, a pair of vortex and anti-vortex, due to superfluid flow past 
an obstacle \cite{Neely}. Furthermore, the recent development to 
observe vortex dipoles in real time \cite{Freilich} opens up the possibility to
examine dynamics. The vortex dipoles can also be considered as the embedded 
section of a vortex ring, which were first observed in BEC as decay product of 
dark solitons \cite{Anderson}. Another phenomenon in superfluids, where
vortex dipoles are crucial is transition to quantum turbulence. The initial 
requirement is the formation of multiple vortex dipoles, followed by
reconnections to form a vortex tangle, which then leads to quantum turbulence
\cite{Seman}. In multi-component condensates, coreless vortices 
\cite{Leanhardt} and Skyrmions \cite{Leslie} have been experimentally observed 
in spinor condensates. Coreless vortices in multi-component spinless 
condensates are, however, yet to be experimentally realized. 

  Theoretically, the flow of the miscible binary and spinor condensates across 
a Gaussian obstacle potential has been investigated
\cite{Susanto,Gladush,Rodrigues}. A related work is the seeding of vortex 
rings when a condensate bubble moves through the bulk of another condensate 
in binary condensates. This is examined in a recent work \cite{Sasaki-11}.  
In addition, phase-separated binary condensates, under suitable conditions, 
are appropriate systems to observe phenomena related to dynamical 
instabilities. These include, modulational instability \cite{Kasamatsu,Ronen},
Kelvin-Helmholtz instability \cite{Takeuchi}, Rayleigh-Taylor 
instability \cite{Gautam,Sasaki}, countersuperflow instability
\cite{Takeuchi-10,Suzuki} and Richtmyer-Meshkov instability \cite{Bezett}. 
Among these, the countersuperflow instability was recently realized 
experimentally \cite{Hamner}. Phase separation, which has been observed 
experimentally \cite{Papp,Tojo}, is also a necessary condition
to observe coreless vortex dipoles with a moving obstacle.

  In this letter we report the theoretical study of vortex dipole formation
when a Gaussian obstacle moves through phase-separated binary condensates.
As a specific case, we consider the binary condensate consisting of 
$^{85}$Rb and $^{87}$Rb. In this, one of the scattering lengths
can be tuned by using magnetic Feshbach resonances \cite{Papp}.
This allows us to choose values of scattering lengths suitable for 
generation of coreless vortex dipoles.


\section{Binary condensates and vortex dipoles}
 Dynamics of weakly interacting binary condensate at zero temperature is well
described by a set of coupled GP equations
\begin{equation}
 \left[ \frac{-\hbar^2}{2m}\nabla^2 + V_i({\mathbf r},t) + 
 \sum_{j=1}^2U_{ij}|\Psi_j({\mathbf r},t)|^2 - 
 i\hbar\frac{\partial}{\partial t}\right]\Psi_i ({\mathbf r},t) = 0
 \label{eq.gp}
\end{equation}
in mean field approximation, where $i = 1, 2$ is the species index. Here 
$U_{ii} = 4\pi\hbar^2a_{ii}/m_i$, where $m_i$ is the mass and $a_{ii}$ is 
the $s$-wave scattering length, is the intra-species interaction, 
$U_{ij}=2\pi\hbar^2a_{ij}/m_{ij}$, where $m_{ij}=m_i m_j/(m_i+m_j)$ is the
reduced mass and $a_{ij}$ is the inter-species scattering length, is the
inter-species interaction, and $V_i({\mathbf r})$ is the trapping potential 
experienced by $i$th species. In the present work, we consider binary 
condensate consisting of $^{85}$Rb and $^{87}$Rb for which $m_1\approx m_2$.

Furthermore, we also consider identical trapping potentials, which are 
axially symmetric, for both the species. The total potential is then 
$$
 V({\mathbf r},t) = \frac{m\omega^2}{2}(x^2 + y^2 + 
                      \beta ^2 z^2) + V_{\rm obs}(x,y,t),
$$
where $V_{\rm obs}(x,y,t) = 
V_0 (t)\exp\lbrace -2([x-x_0(t)]^2+y^2)/w_0^2\rbrace$ is the blue detuned
Gaussian obstacle potential and $\beta$ is the anisotropy parameter. Define 
the oscillator length of the trapping potential
$a_{\rm osc} = \sqrt{\hbar/(m\omega)}$, and consider $\hbar\omega$ as the unit 
of energy. We can then rewrite the equations in dimensionless form with 
transformations
$\tilde{{\mathbf r}} =  \mathbf r/a_{\rm osc}$, 
$\tilde{t} =  t\omega $ and
$\phi_{i}(\tilde{{\mathbf r}},\tilde{t})=   \sqrt{a_{\rm osc}^3/N_i}
\Psi_i({\mathbf r},t)$.
In pancake-shaped traps ($\beta\gg1$), $\phi(\mathbf{r},t)= 
\psi(x,y,t)\zeta(z)\exp({-i\beta t/2})$ \cite{Muruganandam}, where
$\zeta=(\beta/(2\pi))^{1/4}\exp(-\beta z^2/4)$ is the ground state wave 
function in axial direction. The Eq. (\ref{eq.gp}) can then be reduced to the 
two dimensional form 
\begin{eqnarray}
  \left[ -\frac{1}{2}\left(\frac{\partial^2}{\partial x^2} + 
  \frac{\partial^2}{\partial y^2}\right) + \frac{x^2+\alpha_i^2y^2}{2} + 
  V_{\rm obs}(x,y,t) + \right. \nonumber \\
 \sum_{j=1}^2 u_{ij}|\psi_j({\mathbf r},t)|^2 \left. - 
  i\frac{\partial}{\partial t}\right]
  \psi_i ({\mathbf r},t) = 0,
\label{gp_2d}
\end{eqnarray}
where $ u_{ii} = 2 a_{ii}N_i\sqrt{2\pi\beta_i}/a_{\rm osc}$ and 
$ u_{ij} = 2 a_{ij}N_j\sqrt{2\pi\beta_i}/a_{osc}$. Here we have neglected a
constant term corresponding to energy along axial direction as it only shifts
the energies and chemical potentials by a constant number without affecting
the dynamics. In the present work, we
consider $u_{12}>\sqrt{u_{11}u_{22}}$ so that the ground state of the binary 
condensate is phase-separated. Geometry of the density distribution is such
that the species with the lower repulsion energy forms a core and the other
species forms a shell around it. For convenience, we identify the former and 
later as the first and second species, respectively. With this labelling, 
interaction energies $u_{11}<u_{22}$ and for equal populations, this implies 
$a_{11}<a_{22}$.

To be specific, we consider $^{85}$Rb-$^{87}$Rb binary condensate 
with $a_{11} = 460a_0$,  $a_{22} = 99a_0$, and $a_{12} = 214a_0$ as the 
scattering length values and $2 N_1 = N_2 = 10^6$ as the number of atoms. Here 
$a_{11}$ is tunable with magnetic Feshbach resonance \cite{Cornish}. 
With these set of parameters, the stationary state of $^{85}$Rb-$^{87}$Rb 
binary condensate is just phase-separated. The trapping potential and obstacle 
laser potential parameters are same as those considered in Ref. \cite{Neely}, 
i.e. $\omega/(2\pi) = 8$Hz, $\alpha = 1$, $\beta = 11.25$, 
$V_0(0) = 93.0\hbar\omega$, and $w_0 = 10\mu$m. Hereafter we term this set of 
scattering lengths, number of atoms and trapping potential parameters as 
{\em set a}.

   In hydrodynamics, the velocity field of a vortex 
dipole is the vector sum of two component fields. One of the fields arises due
to the inhomogeneous density of the condensate and leads to the precession of an 
off center vortex around the trap center \cite{Jackson-2,Svidzinsky}. In 
addition to this, each vortex has a velocity field which 
varies inversely with the distance from its center, which is experienced by 
the other vortex of vortex dipole. In the present work, we move the obstacle 
along $x$-axis and generate vortex dipoles located symmetrically about 
$x$-axis. If $(x,y)$ and $(x,-y)$ are the locations of the positively and 
negatively charged vortices of the vortex dipole, respectively, then the 
velocity field of the positively charged vortex is
$$
 \mathbf v(x,y) = \omega_{\rm pr}\hat k \times \mathbf r + \frac{1}{2y}\hat i, 
$$
where $\omega_{\rm pr}$ is the rotational frequency of a vortex with charge 
$+1$ in the condensate. A similar equation describes the velocity field of the
negatively charged vortex. 
\begin{figure}
   \includegraphics[width=8.5cm] {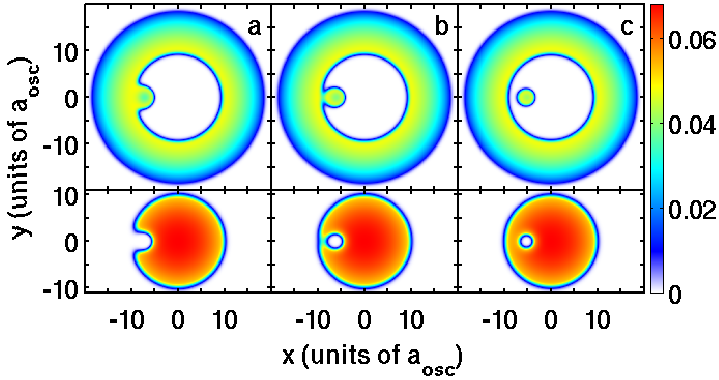}
 \caption{(Color online) Stationary state $|\psi|$ of binary condensate with
           obstacle potential at (a) $-6.0a_{\rm osc}$, (b)$-5.9a_{\rm osc}$
           and (c)$-5.0a_{\rm osc}$.
          }
\label{stat_fig}
\end{figure}


\section{Obstacle modified density}
  To examine the density perturbations from the obstacle beam, 
let $R_{\rm in}$ be the radius of the inner species or the interface boundary. 
And, let $R_{\rm out}$ be the radial extent of the outer species. In the 
absence of the obstacle beam, the chemical potential of first and second 
species in scaled units are 
$\mu_1=R_{\rm in}^2/4 + u_{11}/(\pi  R_{\rm in}^2)$ and $R_{\rm out}^2/2$, 
respectively. The obstacle beam initially ($t=0$) located 
at $(-R_{\rm out},0)$ traverses towards the center with velocity $v_{\rm obs}$ 
and the intensity is ramped down at the rate $-\partial V_0/\partial t=\eta$. 
The location of the beam at a later time is 
$x_0(t) = -R_{\rm out} + v_{\rm ob}t$,
and  intensity of the beam is $V_0(t) = V_0(0) -\eta t$,
where  $V_0(0)$ is the initial intensity of the obstacle beam. At the starting 
point, the total potential $V(R_{\rm out},0,0) > R_{\rm out}^2/2$ and the 
density of the outer species $|\psi_2|^2$ is zero around the center of the 
obstacle beam. However, as it traverses the condensates with decreasing 
intensity, at some later time $t'$, $V(x_0(t'),0,0) < R_{\rm out}^2/2$. 
Density $|\psi_2|^2$ is then finite within the obstacle. For compact notations,
hereafter we drop the explicit notation of time dependence while writing 
$x_0(t)$ and  $V_0(t)$.

 A critical requirement to form coreless vortices is complete immersion  of
the obstacle beam within $n_1$. Based on the previous discussions, as the beam 
approaches the origin, the last point of contact between the beam and interface 
at $R_{\rm in}$ lies along $x$-axis. To determine the condition when complete
immersion occurs, consider the total potential along $x$-axis around the
obstacle potential
\begin{eqnarray}
 V(x,0,t) &\approx & \frac{x^2}{2} + V_0(t) \left[1 -2\frac{(x-x_0(t))^2}{w^2}
                     \right.  \nonumber \\
           && \left .  + 4\frac{(x-x_0(t))^4} {w^4}\right],
\label{V_x0t}
\end{eqnarray}
where, the Gaussian beam potential is considered up to the second order term.
The expression is appropriate in the neighborhood of the beam, and along
$x$-axis it has one local minima ($x_{\rm min}$ ) and maxima each. There is 
also a global minima, however, it is not the correct solution as it lies in 
the domain where $x>w/\sqrt{2}$ and hence outside the domain of validity of
Eq.~(\ref{V_x0t}). Correct global minima is located at 
$x\approx 0$ and is associated with the harmonic potential. The obstacle is 
considered well immersed when $x_{\rm min}$ is located at the interfacial 
radius $R_{\rm in}$, and let $t_{\rm im}$ be the time when it occurs. 

 When the obstacle beam is well inside the inner species, within the obstacle 
beam $n_1$ is zero but $n_2$ is nonzero. It then forms a second interface 
layer, which embeds a bubble of the $n_2$ within $n_1$. Recollect, the first 
interface layer is located at $R_{\rm in}$ and it is where $n_2$ encloses 
$n_1$. The second interface, unlike the one at $R_{\rm in}$, is a 
deformed-ellipse and we label it as $\Gamma$. Around the interface, the two 
condensates mix with a penetration depth 
$$
    \Lambda_i = \xi_i\left [ \frac{\sqrt{a_{11}a_{22}}}
          {a_{12}-\sqrt{a_{11}a_{22}}}\right ] ^{1/2},
$$
and the density of the minority species decays exponentially. 

  The transition from a single continuous interface to two separate boundaries 
at $R_{\rm in}$ and $\Gamma$, when the obstacle crosses $R_{\rm in}$, is 
smooth in TF approximation and that is how we have defined $t_{\rm im}$. There 
are, however, strong perturbations when surface tension is considered, and the 
separation of the two interfaces occurs when the beam is deep inside $n_1$. 
Prior to the separation, the interface is deformed to accommodate a long neck 
region where $n_1$ and $n_2$ are non zero. As the interface splits into two, 
there are large deformations from the equilibrium interface geometry, and 
surface tension generates a restoring force to bring it to equilibrium 
geometry. This creates density patterns with high curvature and initiates 
formation of the coreless vortex dipoles.

\section{Obstacle assisted bubble}
 At a time $\Delta t$ after the obstacle is immersed in $n_1$, the location 
and amplitude of the obstacle potential are
\begin{eqnarray}
  x_0(t_{\rm im}+\Delta t) & = & -R_{\rm out} + v_{\rm ob}\times(t_{\rm im} 
                               + \Delta t),  \nonumber \\
  V_0(t_{\rm im}+\Delta t) & = & V_0(0) - \eta\times(t_{\rm im} + \Delta t).
                   \nonumber
\end{eqnarray}
Equilibrium TF $n_2$ within the obstacle potential at this instant of time is
$$
  n_{2\Gamma} (x, y, t_{\rm im} + \Delta t ) = \frac{\mu_2 
          - V(x, y, t_{\rm im} + \Delta t)}{u_{22}}.
 \label{n2_den}
$$
This, however, is higher than the density distribution at $t_{\rm im}$, that is
$n_{2\Gamma} (x, y, t_{\rm im} + \Delta t )> n_{2\Gamma} (x, y, t_{\rm im})$ as
the potential $ V $ is lower. This is on account of two factors: first,
the amplitude of the obstacle potential decreases with time; and second, the
harmonic oscillator potential is lower at $x_0(t_{\rm im}+\Delta t)$. The
number of atoms, however, does not change from the value at $t_{\rm im}$ 
unless there is a strong Josephson current. Density $n_{2\Gamma}$  is thus
below the equilibrium value once the obstacle beam is well within $n_1$. 
This creates a stable bubble of $n_2$ assisted or trapped within the beam
and is transported through the $n_1$.

 Departure of $n_2$ from the equilibrium is not the only density evolution
within the beam. There is a progressive change of $n_{1\Gamma} $ (density
of first species within the obstacle beam) as the beam moves deeper into 
$n_1$. At time $t_{\rm im}$, when the obstacle is completely immersed in 
$n_1$ the effective potential, experienced by $n_1$, 
$V(x, y, t_{\rm im}) + n_{2\Gamma}u_{12}$ is larger than $\mu_1$. So, $n_1$ is
zero within the beam. However, if the rate of ramping $\eta$ is such that at 
a later time $V(x, y, t_{\rm im} + \Delta t) + n_{2\Gamma}u_{12}< \mu_1$, 
while the beam is still within $n_1$, there is a finite $n_1$ within the beam.
Since $a_{12}>\sqrt{a_{11}a_{22}}$ for the condensate, in TF approximation 
the bulk values of $n_{1\Gamma}$ and $n_{2\Gamma}$ can not be simultaneously 
non-zero. At the same time, $n_{2\Gamma}$ is forbidden to migrate to the 
bulk $n_2$ due to the $n_1$ generated potential barrier in the region between 
interfaces $\Gamma$ and $R_{\rm in}$. To accommodate both $n_1$ and $n_2$ 
within the beam, the shape of interface $\Gamma$ is transformed to increase 
$n_{2\Gamma}$. So that $n_2$ is zero in certain regions within the beam where 
the condition, $V(x, y, t_{\rm im} + \Delta t) + n_{2\Gamma}u_{12}<\mu_1$, 
is satisfied. This mechanism is responsible for obstacle 
assisted transport of $n_2$ across $n_1$. 

\begin{figure}
   \includegraphics[width=8.5cm] {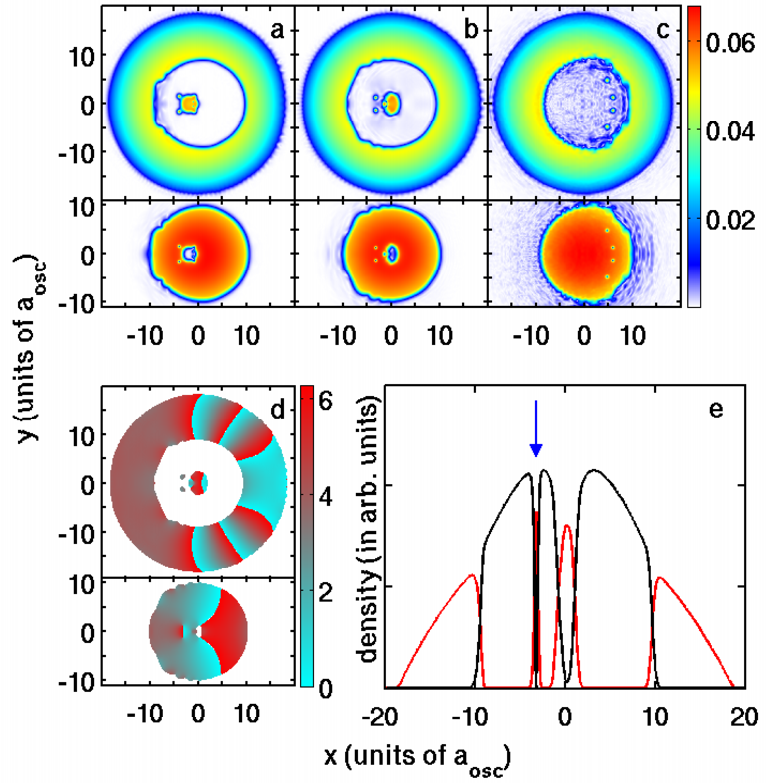}
 \caption{(Color online) $|\psi_i|$ and phase of binary condensates after the 
          creation of coreless vortex dipole. (a) Coreless vortex dipole is 
          fully formed but yet to dissociate from the obstacle, (b) center of 
          obstacle beam is at origin and coreless vortex dipole is separated, 
          (c) additional coreless vortex dipoles are generated when the obstacle 
          reaches the interface, (d) phase of the binary condensate 
           corresponding to (b), and (e) densities of the condensates parallel 
          to $x$-axis and passing through the center of the coreless vortex. Blue
          arrow marks the center of the coreless vortex dipole. 
          }
\label{vortex_fig}
\end{figure}


\section{Energetic stability of normal versus coreless vortex dipoles}

 We use TF approximation to compare the energetic stabilities of normal
and coreless vortex dipoles.

\subsubsection{Normal vortex dipole}
Assuming that the vortex affects the density of the condensate only  within the core
regions, we can adopt the following {\em ansatz} for binary condensate with a
normal vortex dipole at $(v_1,\pm v_2)$
\begin{eqnarray}
 \psi_1(r)  & = & \left \{ 
     \begin{aligned}
        & 0 \!&& x^2+y^2 > R_{\rm in}^2\\
        & 0 \!&& [(x-v_1)^2+(y\pm v_2)^2] \leqslant \xi^2\\
        & \sqrt{\frac{\mu_1 - V(x,y)}{u_{11}}} 
              && \left \{ 
                   \begin{aligned}
                     & x^2+y^2 \leqslant R_{\rm in}^2~\&\\
                     &~[(x-v_1)^2+(y\pm v_2)^2] > \xi^2
                  \end{aligned} \right .
     \end{aligned} \right .   \\
 \psi_2(r)  &=&  \left \{ 
     \begin{aligned}
        & \sqrt{\frac{\mu _2-V(x,y)}{u_{22}}}
                 &&R_{\rm in}^2\leqslant (x^2+y^2) \leqslant R_{\rm out}^2\\
        & 0&&(x^2+y^2) > R_{\rm out}^2\\
        & 0&&(x^2+y^2) < R_{\rm in}^2.
     \end{aligned} \right .
\end{eqnarray}
The vortex dipole contributes mainly through the kinetic energy of $\psi_1$,
which may be approximated with the  value of single species condensate given
in Ref.\cite{Zhou}
\begin{equation}
   E_{\rm vd} = \frac{2\mu _1}{u_{11}}\ln\left(\frac{2v_2}{\xi}\right),
\end{equation}
where $\xi = 1/\sqrt{2\mu_1}$ is the coherence length of inner species.
Using these {\em ansatz} the number of atoms are
\begin{eqnarray}
N_1 & = & \frac{\pi  \left(1+4 v_1^2 \mu _1+4 v_2^2 \mu _1-8 
          \mu _1^2-2 R_{\rm in}^4 \mu _1^2+8 R_{\rm in}^2 
          \mu _1^3\right)}{8 u_{11} \mu _1^2},\nonumber \\
N_2 & = & \frac{\pi  \left(R_{\rm in}^2-2 \mu _2\right){}^2}{4 u_{22}}.
\end{eqnarray}
In a similar way, we can evaluate the energy of the entire condensate.
\subsubsection{Coreless vortex dipole}
\begin{center}
\begin{figure}[ht]
\includegraphics[width=8.5cm] {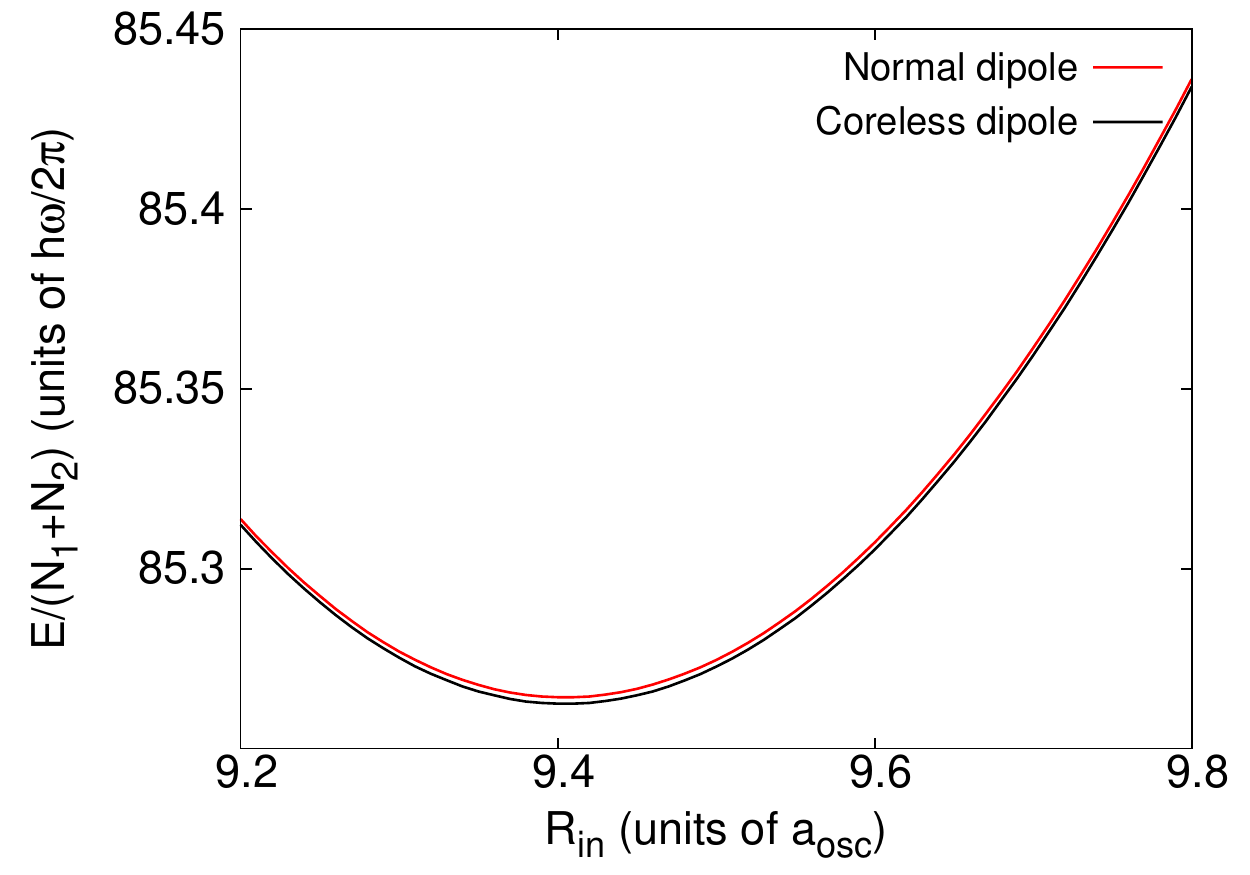}
\caption{(Color online) The energy of the binary condensate with $V_0=0$ and rest of the
parameters same as those in the parameters {\em set a} as a function of
$R_{\rm in}$. The condensate has a vortex dipole located at $(0,\pm 1)$.
Black and blue curves are for coreless and normal vortex dipoles
respectively.}
\label{fig_coreless_vortex_dipole1}
\end{figure}
\end{center}

\begin{figure}[ht]
\includegraphics[width=8.5cm] {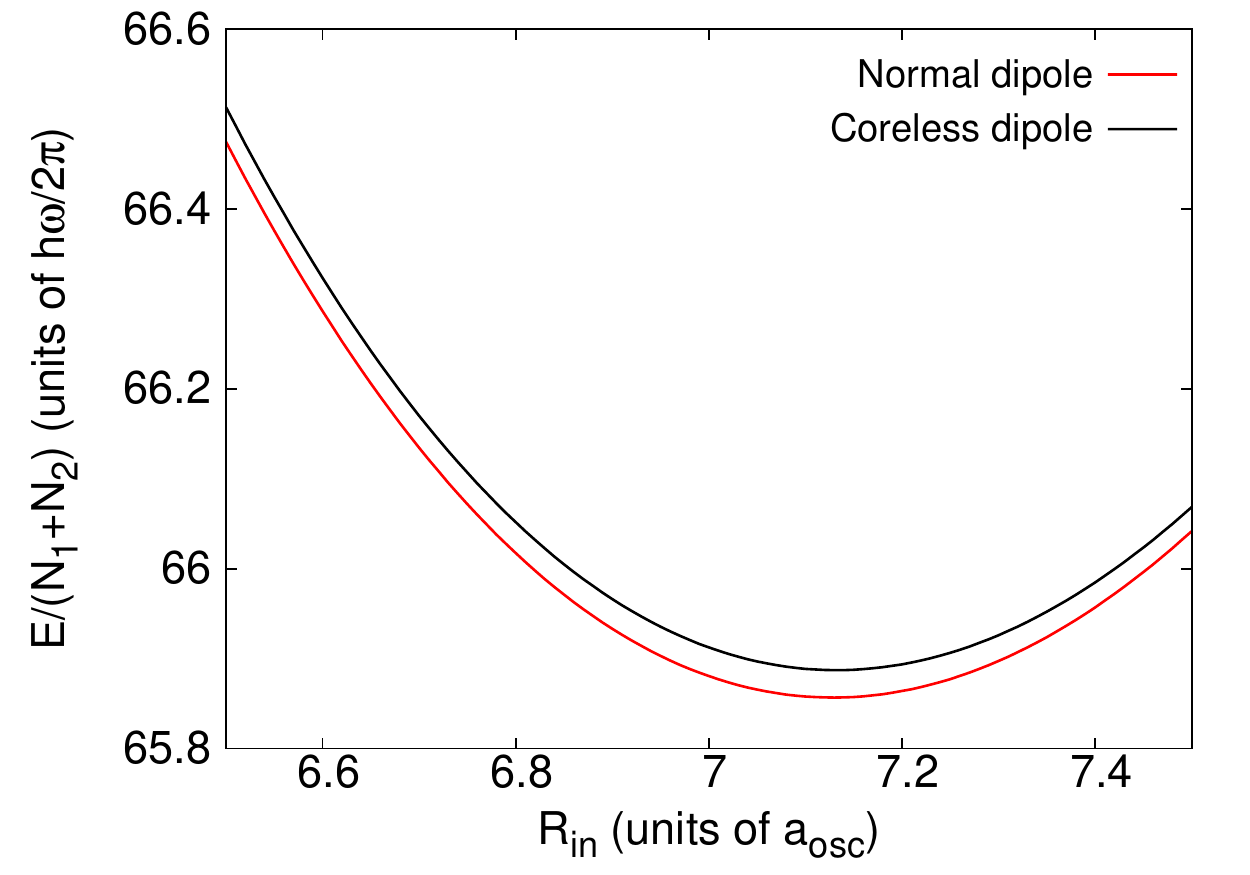}
\caption{(Color online) The energy of the binary condensate with $N_1=N_2=10^6$,
$a_{11}=51a_0$, $a_{22}=99a_0$, $V_0 = 0$, and rest of the parameters same as
those in parameters {\em set a} as a function of $R_{\rm in}$. The condensate
has vortex dipole located at $(0,\pm 1)$. Black and blue curves are for
coreless and normal vortex dipoles respectively.}
\label{fig_coreless_vortex_dipole2}
\end{figure}

For coreless vortex dipole, we adopt the {\em ansatz}
\begin{eqnarray}
 \psi_1(r)  & = &  \left \{
      \begin{aligned}
         & 0 && x^2+y^2 > R_{\rm in}^2\\
         & 0 && [(x-v_1)^2+(y\pm v_2)^2] \leqslant \xi^2\\
         & \sqrt{\frac{\mu_1 - V(x,y)}{u_{11}}} 
              && \left \{ 
                \begin{aligned}
                   & x^2+y^2 \leqslant R_{\rm in}^2~\&\\
                   &~[(x-v_1)^2+(y\pm v_2)^2] > \xi^2,
                \end{aligned} \right .
      \end{aligned} \right . \\
 \psi_2(r)  & = & \left \{
      \begin{aligned}
         & \sqrt{\frac{\mu _2-V(x,y)}{u_{22}}}
                &&\left \{
                   \begin{aligned}
                      & R_{\rm in}^2\leqslant (x^2+y^2) \leqslant R_{\rm out}^2~||~\\
                      & [(x-v_1)^2+(y\pm v_2)^2] \leqslant \xi^2
                   \end{aligned} \right . \\
         & 0&&(x^2+y^2) > R_{\rm out}^2\\
         & 0&& \left \{
                \begin{aligned}
                   & x^2+y^2 < R_{\rm in}^2~\&\\
                   &~[(x-v_1)^2+(y\pm v_2)^2] > \xi^2.
                \end{aligned} \right .
      \end{aligned} \right .
\end{eqnarray}
Using these {\em ansatz} the modified expressions for $N_2$ is
\begin{eqnarray}
N_2 &=& \frac{\pi}{8 u_{22} \mu _1^2}\left(2 R_{\rm in}^4 \mu _1^2
       + 8 \mu _1 \mu _2-8 R_{\rm in}^2 \mu _1^2 \mu _2+8 \mu _1^2 \mu _2^2 -1 
         \right.       \nonumber \\
    & &\left . -4 v_1^2 \mu _1-4 v_2^2 \mu _1\right).
\end{eqnarray}
As done earlier, we can also calculate the total energy $ E$ of the system. The important
change in $E$ is the inclusion of interface interaction energy $E_{\rm int}$.
It arises from the interface interactions at the cores of the vortex and
antivortex. Based on Ref.~\cite{Timmermans},
\begin{equation}
 E_{\rm int} = \frac{8}{3}Pb\pi\xi\left (\frac{a_{12}}{\sqrt{a_{11}a_{22}}}-1
               \right ),
\end{equation}
where $P$ is the pressure on the circumference of the cores and
\begin{equation}
   b = 2\left [ \frac{3(\mu_1+\mu_2)\sqrt{a_{11}a_{22}}}{4\mu_1\mu_2
          (a_{12}-\sqrt{a_{11}a_{22}})} \right ] ^{1/2}.
\end{equation}
In both the case, i.e. with normal and coreless vortex dipoles, the 
energy can be minimized with the constraint of the fixed number of atoms 
and $R_{\rm in}$ as a minimization parameter. For the parameters {\em set a}
without obstacle potential, the coreless vortex dipole has lower energy than
the normal vortex dipole and is shown in
Fig.~\ref{fig_coreless_vortex_dipole1} for the vortex dipole located
at $(0,\pm1)$.
 For $N_1=N_2=10^6$, $a_{11}=51a_0$, $a_{22} = 99a_0$ and rest
of the parameters same as in parameter {\em set a}, condensate with the normal
vortex dipole has lower energy than the one with coreless vortex dipole
(see Fig.~\ref{fig_coreless_vortex_dipole2}). These results are in very good
agreement with the numerical results.


\section{Numerical results and conclusions}To examine the formation of coreless vortex dipoles in finer
detail, we resort to numerical solution of Eq. (\ref{gp_2d}) with a modified 
version of the split-step Crank-Nicholson code reported in 
Ref. \cite{Muruganandam}. Consider obstacle potential is initially 
located in the outer component and is moved across the interface, towards
the origin. For this case, we consider $^{85}$Rb-$^{87}$Rb binary condensate 
with parameters {\em set a}, however, with maximum value of obstacle laser
potential $V_0(0) = 125.0$. The obstacle potential is initially located at 
$x = -15 a_{\rm osc}$. The obstacle moves with the speed of $180\mu$m/s,
progressively decreases in strength with rate constant $\eta = 10.1$ 
(in scaled units), and vanishes at $x = 8a_{\rm osc}$. The obstacle potential 
creates a normal vortex dipole as it traverses $n_2$. As the obstacle 
penetrates the interface, it carries the vortex dipole generated in the outer 
component in its region of influence. Further motion of the obstacle, in 
$n_1$, creates coreless vortices.

 The key factor which influences the generation of coreless vortex dipoles is 
the deformation at the aft region of the obstacle confined $n_2$. The
deformation accompanied by large mixing is initiated when the interface 
is about to break up. This is evident even in the stationary state density 
distribution shown in Fig. \ref{stat_fig}(b). At break up, the interface 
repulsion and potential gradient are highest along the $x$-axis and lead to 
the formation of a dimple. Curvature is large around the deformed interface, 
and flow of $n_1$ past it generates vortex dipoles. However,
as the vortex dipoles are generated within the penetration zone, the build up 
of $n_1$ around the vortex core drives $n_2$ from the penetration zone to the 
core of the vortex. For the parameters considered, first coreless vortex 
dipole is formed soon after the interface break up and is shown in 
Fig. \ref{vortex_fig}(a) at $t=0.21s$. The vortex dipole is almost detached 
when the obstacle reaches origin, shown in Fig. \ref{vortex_fig}(b). From the 
phase, Fig. \ref{vortex_fig}(d), it is evident that the phase singularity 
is associated with $n_1$ and $n_2$ is non-zero at the core. This is seen in 
the plot of densities along a line parallel to $x$-axis and passing through the 
vortex core, Fig. \ref{vortex_fig}(e). In the figure, the blue arrow marks the 
location of the vortex core. Another important density modification is, 
although the obstacle potential is repulsive, $n_2$ has a maxima at the 
center. This is due to the repulsion energy from $n_1$. More coreless vortex 
dipoles are created when the obstacle crosses the center of harmonic potential, 
\ref{vortex_fig}(c).  

\begin{center}
\begin{figure}
\includegraphics[width=8.5cm] {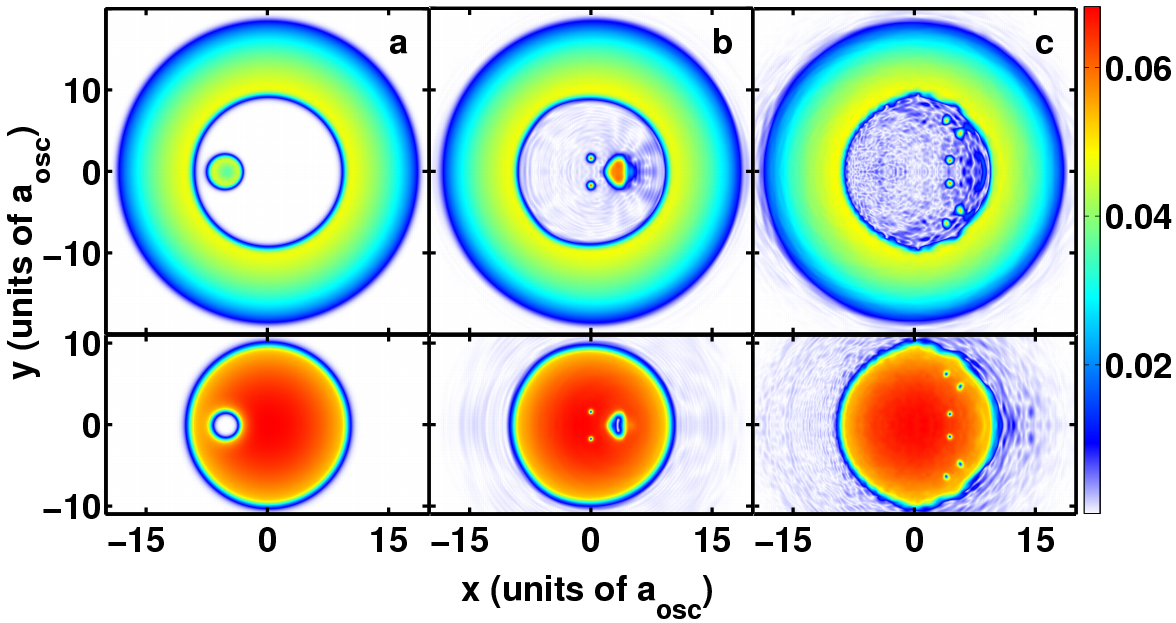}
\caption{(Color online) The generation of triply charged vortex dipole in the
         $^{85}$Rb-$^{87}$Rb binary condensate with parameters {\em set a}.
         The obstacle potential, initially located at $x = -5.0 a_{\rm osc}$,
         is moved with a velocity of $220\mu$m/s up to $x = 5 a_{\rm osc}$.
         First, second, and third columns are the solutions at $t=0$s,
         $t=0.14$s and $t = 0.32$s, respectively.}
\label{triply_charged_vdipole}
\end{figure}
\end{center}

 The other initial configuration is to place the obstacle potential within
$n_1$. If $V_{\rm obs}$ is such $\mu_1 < V(\mathbf r, t_0) < \mu_2$, then 
within the obstacle potential $n_2$ is nonzero. Initial density distribution 
is like in Figs. \ref{stat_fig}(c) and \ref{triply_charged_vdipole}(a).
Due to inertia, $n_2$ lags behind the beam
when the obstacle suddenly starts to move. When the obstacle has shifted from
its initial position $x_0$ to $x_0 + \delta x_0$, the points of intersection of
the inner interfaces (assumed circular with radii $R_{\Gamma}$ and centered around
$x_0$ and $x_0 + \delta x_0$) are
\begin{equation}
(x_c,y_c) = \left(\frac{1}{2} (2 x_0+\text{$\delta $x}),
            \pm \frac{1}{2} \sqrt{4 {R_{\Gamma}}^2-
            \text{$\delta $x}^2}\right).
\label{cross_over}
\end{equation}
Due to inertia $n_2$ still occupies the region $\delta \Gamma$ defined as 
$$
(x-x_0)^2+y^2  <  R_{\Gamma}^2<(x-x_0-\delta x)^2+y^2
$$
The potential experienced by $n_2$ along the left interface of  $\delta \Gamma$
decreases as one moves from $x$-axis. This leads to redistribution of
$n_2$ in $\delta \Gamma$ region and creation of a pressure difference 
$\delta P = (n_1^2u_{11}-n_2^2u_{22})/2\geqslant0$ at the left bounding arc
of $\delta\Gamma$. Along this arc, $\delta P$ decreases from 
$(x_0-R_{\Gamma},0)$ to $(x_c,y_c)$ and tends to flatten it. Another equally
important dynamical process is the redistribution of $n_1$ as
$V_{\rm eff} = V(x,y) + n_2(x,y)u_{12}$ increases along the left interface 
from $x$-axis to $(x_c,y_c)$. Due to this, $n_1$ start to penetrate the
interface from the point where $V_{\rm eff} \leqslant \mu_1$. Thus the 
repulsive interaction at the interface and gradient of harmonic potential, 
combine to form a dimple. The dimple formation initiates the formation of 
(coreless) ghost vortices in the obstacle region \cite{Fujimoto}, which 
detach to form coreless vortex dipoles as is shown in 
Figs.~\ref{triply_charged_vdipole}(b)-(c).

We have studied the motion of the Gaussian obstacle across a phase-separated
binary condensate. With the possibility of tuning one of the scattering lengths
using Feshbach resonances, these condensates can be used to experimentally 
realise the obstacle assisted transport of one species across another as well 
as coreless vortices. Using both TF approximation and exact numerical 
solutions of coupled GP equations, we have shown that coreless vortex dipoles
can be energetically more preferable than normal vortex dipoles.


\begin{acknowledgements}
We thank S. A. Silotri, B. K. Mani, and S. Chattopadhyay for very useful 
discussions. The numerical computations reported in the paper were done on
the 3 TFLOPs cluster at PRL. The work of PM forms a part of Department of 
Science and Technology (DST), Government of India sponsored research project.

\end{acknowledgements}



\begin{thebibliography}{99}
  \bibitem{Neely}
          T.~W.~Neely, E.~C. Samson, A.~S.~Bradley, M.~J.~Davis, and 
          B.~P.~Anderson,
          Phys. Rev. Lett. {\bf 104}, 160401 (2010).
  \bibitem{Freilich}
          D.~V.~Freilich, D.~M.~Bianchi, A.~M.~Kaufman, T.~K.~Langin, and 
          D.~S.~Hall,
          Science  {\bf 329}, 1182 (2010).
  \bibitem{Anderson}
          B.~P.~Anderson, P.~C.~Haljan, C.~A. Regal, D.~L.~Feder, 
          L.~A.~Collins, C.~W.~Clark, and E.~A.~Cornell,
          Phys. Rev. Lett. {\bf 86}, 2926 (2001).
  \bibitem{Seman}
          J.~A.~Seman, {\em et al.},
          arXiv:1007.4953v4.
  \bibitem{Leanhardt}
          A.~E.~Leanhardt, Y.~Shin, D.~Kielpinski, D.~E.~Pritchard, and 
          W.~Ketterle,
          Phys. Rev. Lett. {\bf 90}, 140403 (2003).
  \bibitem{Leslie}
          L.~S.~Leslie, A.~Hansen, K.~C.~Wright, B.~M.~Deutsch, and 
          N.~P.~Bigelow,
          Phys. Rev. Lett. {\bf 103}, 250401 (2009)
  \bibitem{Susanto}
          H.~Susanto, P.~G.~Kevrekidis, R.~Carretero-Gonzalez, B.~A.~Malomed, 
          D.~J.~Frantzeskakis, and A.~R.~Bishop,
          Phys. Rev. A {\bf 75}, 055601 (2007).
  \bibitem{Gladush}
          Yu.~G.~Gladush, A.~M.~Kamchatnov, Z.~Shi, P.~G.~Kevrekidis, 
          D.~J.~Frantzeskakis, and B.~A.~Malomed,
          Phys. Rev. A {\bf 79}, 033623 (2009).
  \bibitem{Rodrigues}
          A.~S.~Rodrigues, {\em et al.},
          Phys. Rev. A {\bf 79}, 043603 (2009).
  \bibitem{Sasaki-11}
          K.~Sasaki, N.~Suzuki, and H.~Saito,
          Phys. Rev. A {\bf 83}, 033602 (2011).
  \bibitem{Kasamatsu}
          K.~Kasamatsu, and M.~Tsubota, 
          Phys. Rev. Lett. {\bf 93}, 100402 (2004).
  \bibitem{Ronen}
          S.~Ronen, J.~L.~Bohn, L.~E.~Halmo, and M.~Edwards, 
          Phys. Rev.  A {\bf 78}, 053613 (2008).
  \bibitem{Takeuchi}
          H.~Takeuchi, N.~Suzuki, K.~Kasamatsu, H.~Saito, and M.~Tsubota,
          Phys. Rev. B {\bf 81}, 094517 (2010).
  \bibitem{Gautam}
          S.~Gautam, and D.~Angom, 
          Phys. Rev. A {\bf 81}, 053616 (2010).
  \bibitem{Sasaki}
          K.~Sasaki, N.~Suzuki, D.~Akamatsu, and H.~Saito, 
          Phys. Rev. A {\bf 80}, 063611 (2009).
  \bibitem{Takeuchi-10}
          H.~Takeuchi, S.~Ishino, and M.~Tsubota,
          Phys. Rev. Lett. {\bf 105}, 205301 (2010).
  \bibitem{Suzuki}
          N.~Suzuki, H.~ Takeuchi, K.~Kasamatsu, M.~Tsubota, and H.~Saito,
          Phys. Rev. A. {\bf 82}, 063604 (2010).
  \bibitem{Bezett}
          A.~Bezett, V.~Bychkov, E.~Lundh, D.~Kobyakov, and M.~Marklund,
          Phys. Rev. A. {\bf 82}, 043608 (2010).
  \bibitem{Hamner}
          C.~Hamner, J.~J.~Chang, P.~Engels, and M.~A.~Hoefer, 
          Phys. Rev. Lett. {\bf 106}, 065302 (2011).
  \bibitem{Papp}
          S.~B.~Papp, J.~M.~Pino, and C.~E.~Wieman,
          Phys. Rev. Lett. {\bf 101}, 040402 (2008).
  \bibitem{Tojo}
          S.~Tojo, Y.~Taguchi, Y.~Masuyama, T.~Hayashi, H.~Saito, and T.~Hirano,
          Phys. Rev. A {\bf 82}, 033609 (2010).
  \bibitem{Muruganandam}
          P.~Muruganandam, and S.~K.~Adhikari,
          Comp. Phys. Comm. {\bf 180}, 1888 (2009).
  \bibitem{Cornish}
          S.~L.~Cornish, N.~R.~Claussen, J.~L.~Roberts, E.~A.~Cornell, and 
          C. E. Wieman,
          Phys. Rev. Lett. {\bf 85}, 1795 (2000).  
  \bibitem{Svidzinsky}
          A.~A.~Svidzinsky and A.~L.~Fetter,
          Phys. Rev. Lett. {\bf 84}, 5919 (2000).
  \bibitem{Jackson-2}
          B. Jackson, J.~.F.~McCann, and C.~.S.~Adams
          Phys. Rev. A {\bf 61}, 013604 (1999).
  \bibitem{Zhou}
          Q.~Zhou and H.~Zhai,
          Phys. Rev. A {\bf 70}, 043619 (2004).
  \bibitem{Timmermans}
          E.~Timmermans,
          Phys. Rev. Lett. {\bf 81}, 5718 (1998).
  \bibitem{Fujimoto}
          K.~Fujimoto and M.~Tsubota,
          Phys. Rev. A {\bf 83}, 053609 (2011).
\end{thebibliography}
\end{document}